# Towards realistic modeling of IP-level routing topology dynamics

Clémence Magnien, Amélie Medem and Fabien Tarissan

UPMC – Sorbonne Universités
CNRS
LIP6 Laboratory
{firstname.lastname}@lip6.fr

**Abstract.** Many works have studied the Internet topology, but few have investigated the question of how it evolves over time. This paper focuses on the Internet routing IP-level topology and proposes a first step towards realistic modeling of its dynamics. We study periodic measurements of routing trees from a single monitor to a fixed destination set and identify invariant properties of its dynamics. We then propose a simple model for the underlying mechanisms of the topology dynamics. Simulations show that it effectively captures the observed behaviors, thus providing key insights of relevant mechanisms governing the Internet routing dynamics.

## 1 Introduction

Studying the structure of the Internet topology is an important and difficult question. No official map being available, researchers have to conduct costly measurement campaigns, and deal with the fact that the obtained data can be biased [**?**,**?**]. Studying the dynamics of this topology is therefore an equally hard, if not harder, problem.

In this paper, we use an orthogonal approach to obtain insight on the dynamics of the routing topology observed at the IP-level. We study *ego-centered views* of this topology. Given a monitor and a fixed set of destinations, one such view is obtained by measuring the routes from the monitor to the destinations. This can be performed quickly and with low network load with the `tracetree` tool [1]. Repeating this measurement periodically therefore allows to study the dynamics of this view.

Previous work has shown that ego-centered views exhibit strong dynamics, and in particular that the set of observed nodes evolves much more quickly than what was previously expected [2]. Here, we analyze in depth this dynamics (Section 2), and find that two factors play a key role in the observed dynamics: load-balancing routers, and the evolution of the routing topology (Section 3).

Based on these observations, we propose in Section 4 a baseline model for the routing dynamics in the Internet that incorporates routing modifications and load balancing. We use the most simple choices for modeling these factors, and show in Section 5 that this model is able to accurately reproduce the behaviors observed in real data. This shows that simple mechanisms such as the ones we take into account play a key role in the Internet routing topology dynamics, giving a strong explanatory value to our model. As such, it represents a significant first step towards the modeling of the Internet IP-level topology and its dynamics.

## 2 Dynamics analysis

The `tracetree` tool [1] collects the *ego-centered view* from a given monitor to a given set of destinations by measuring the routes from this monitor to each destination. This corresponds to a subset of the routing topology, in which nodes are the IP-addresses of routers, and a link exists between two nodes if the corresponding routers are connected at the IP level. Note that the routing topology is different from the physical topology, as two routers may be physically connected by a link that is not used for routing. Note also that we only observe a subset of the whole routing topology, as measuring the routes from a single monitor to a limited set of destinations certainly does not allow to discover all nodes and links in this topology. Moreover, this subset is not representative of the whole topology, see for instance [**?**,**?**]. Keeping this in mind, we will see that we are still able to make interesting observations about the *dynamics* of this topology.

Running the `tracetree` tool periodically allows to capture the dynamics of ego-centered views. We collected two datasets in this way. The first one, `woolthorpe`, was collected from a monitor in University Pierre and Marie Curie in Paris towards a set of $3,000$ destinations. The collection frequency is of one measurement round every 15 minutes approximately. It started in  December, 2010 and ended in  June, 2011, which represents a total of 17,450 rounds. The second one, `ovh`, was collected from a French server hosting company. Only $500$ destinations were used in order to increase the measurement frequency, which is of one round every one and a half minute approximately. It started in October, 2010 and ended in September, 2011, which represents a total of $318,000$ rounds. In both cases, the destinations were chosen by sampling random IP addresses that answered to a ping at the time of the list creation [1]. These datasets are publicly available [3].

It is possible that, at a given time, several routes to a same destination co-exist, in particular because of load balancing. Therefore, two consecutive measurement rounds may capture different routes to a same destination even if no routing change has happened. We study this in the next section, and present below the main characteristics of the *observed* dynamics. Notice that previously measured datasets are available, for different durations, at different times since 2008 [3]. We performed our analysis on a representative set of these datasets, and made similar observations to the ones we present here. This shows the generality of our observations.

*Discovery of new* IP *addresses.* A previous study of the same type of data has shown that these measurements continuously discover new IP addresses that had never been

[1] Previous work has indeed shown that tracing routes to unused IP addresses can introduce measurement artifacts [**?**].

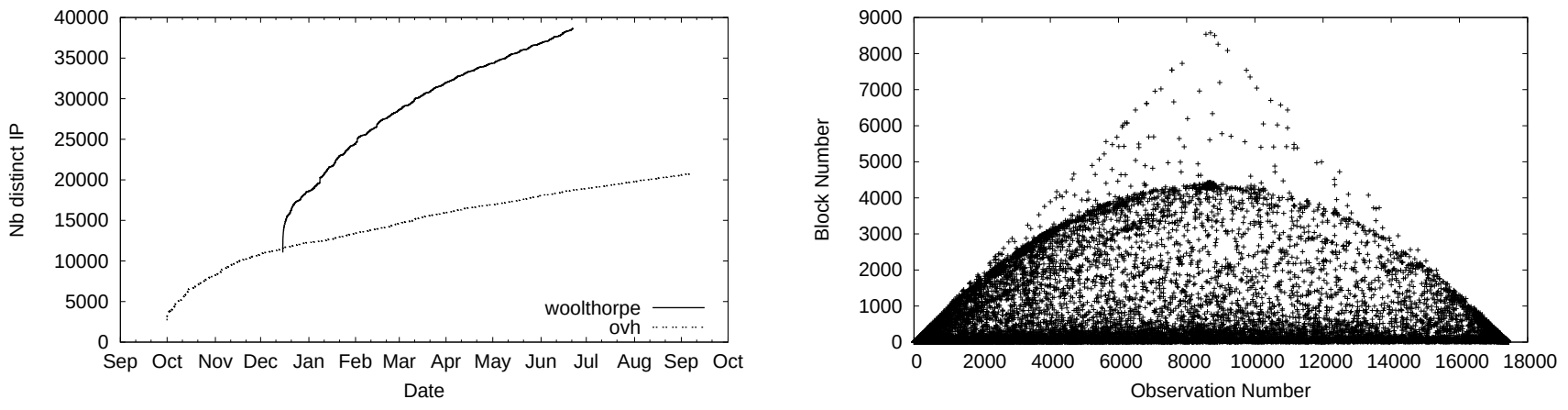

(a) Number of distinct IP addresses observed since measurement beginning.

(b) Observation number vs. block number.

**Fig. 1.** Properties of the observed dynamics.

observed before, at a significant rate [2]. These observations were made on two-months-long measurements. Fig. 1(a) shows that it is also true for very long measurements. It presents the number of IP addresses observed since the beginning of the measurement, for both datasets. A dot $(x, y)$ in this figure means that $y$ different addresses have been observed at least once before time $x$. We see that, after an initial fast growth, the plot increases significantly for extended periods of time.

This plot presents the number of distinct IP addresses observed, and not the number of distinct routers, as in general a router corresponds to several IP addresses, or interfaces. Previous work has however studied the number of distinct ASes discovered by such measurements, and showed that it also increases significantly [2]. This is a good evidence that new routers are actually discovered at a significant rate, even if part of the observed growth may be caused by discovering new interfaces for already observed routers. As there is no method that allows to know with certainty which interfaces correspond to a same router, we limit ourselves to the study of interfaces.

*Stability of* IP *addresses.* To analyze more in depth the dynamics of the ego-centered views, we compute two quantities for each IP address. Its *observation number* is simply the number of distinct rounds it was observed in. An IP address is in general observed in blocks of several consecutive rounds, preceded and followed by one or more rounds during which it is not observed. More precisely, the *block number* of an IP address is the number of groups of consecutive rounds in which it is observed. For example, an IP address which was observed on rounds 1, 3, 4, 7, 8, 9, and 10 has an observation number of 7 and a block number of 3.

Fig. 1(b) presents the correlation between these quantities for the `woolthorpe` dataset. Each dot corresponds to an IP address, and its coordinates are its observation number on the $x$-axis and its block number on the $y$-axis. The plot presents a clear geometric shape: in particular we observe a triangle and a parabola. This can be explained in the following way. The block number cannot be larger than the observation number (therefore $y \leq x$); conversely, the block number cannot be larger than the number of rounds in which the corresponding address was not observed (therefore $y \leq r - x$, $r$ being the total number of measurement rounds). This defines the borders of the triangle.

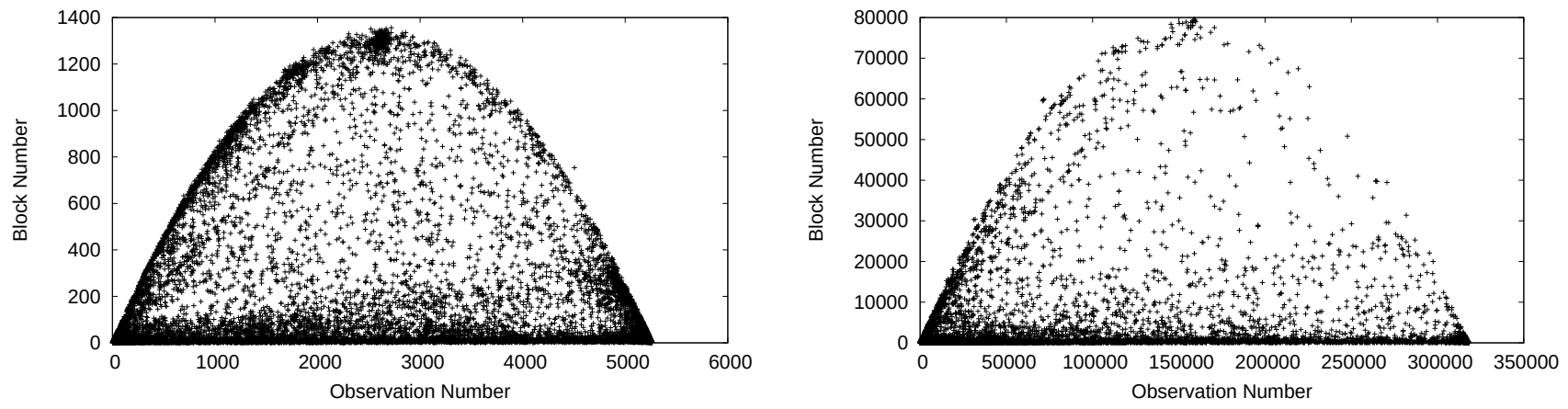

(a) For measurements in which destinations are considered in a random order at each round.

(b) For the *ovh* dataset.

**Fig. 2.** Observation number vs. block number .

The presence a large number of IP addresses close to the parabola can be explained by load-balancing routers. If a load-balancing router randomly spreads traffic among $k$ paths[2], each router belonging to any of these paths has a probability $p = 1/k$ of being observed at each round, leading to an observation number equal to $rp$ approximately. A given round is then the first of a consecutive block of observations for one of these routers with the probability $p$ that this router was observed in this round, multiplied by the probability $1 - p$ that it was not observed in the previous round. Multiplying this probability by $r$ gives the expected block number, which is then equal to $rp(1 - p)$ and is the equation of the parabola. This is a simplification of the real case in which a router may belong to paths used by several load balancers, themselves belonging to paths used by other load balancers. In practice, an IP address belonging to load-balanced paths can have any probability $p$, $0 < p < 1$, of being observed.

IP addresses above the parabola tend to blink more than expected: those which are at the tip of the triangle are observed exactly every other round. Investigation showed that this kind of behavior is caused by the `tracetree` tool design in presence of per-destination load balancing. Faced with two load-balanced routes to destinations $d_1$ and $d_2$, the tool indeed tends to alternate between measuring the route to $d_1$ at one round, then the route to $d_2$ at the next round, and so on.

To confirm this, we performed additional, similar but shorter measurements, in which the tool considers the destination in a random order at each round. The result is presented in Fig. 2(a). It has the same characteristics as the one presented if Fig. 1(b), except there are no points significantly above the parabola, thus confirming that such points are due to the tool itself.

We can also observe a large number of dots close to the $y = x/2$ line. They correspond to addresses that are observed only during a finite part of the measurement, and have during that time a probability $p = 1/2$ of being observed, due to load balancing. If such an IP address is observed with a probability $1/2$ during $k$ rounds, its

[2] It has been shown [15] that per-packet or per-flow load-balancing routers spread `traceroute` probes equally among all paths to the destination, which is roughly equivalent to randomly choosing a path.

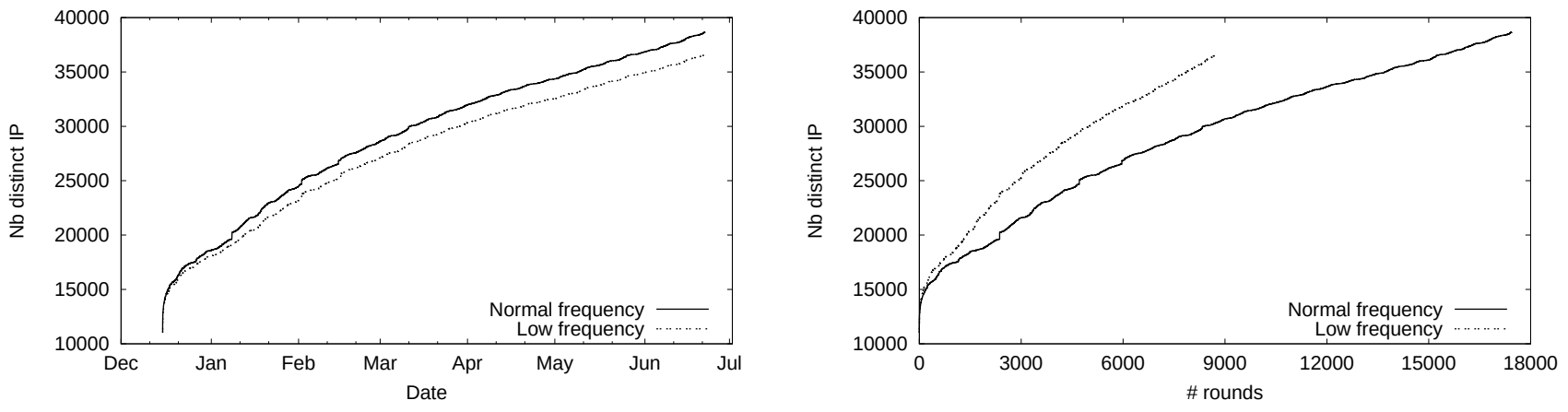

(a) Number of discovered IP addresses as a function of time (in months).

(b) Number of discovered IP addresses as a function of the number of measurement rounds.

**Fig. 3.** How the frequency impacts the number of discovered IP addresses.

observation number will indeed be $x = k/2$, and its expected block number will be $y = k(1/2)^2 = x/2$.

Finally, a large number of IP addresses are close to the $x$-axis. This means that, whether they are observed in a large or small number of rounds, they are mainly observed during blocks of consecutive rounds, with few interruptions.

Figure 2(b) presents the corresponding plot for the `ovh` dataset. It has the same shape than the one in Figure 1(b) for the `woolthorpe` dataset, except that there are no points significantly above the parabola. Manual inspection indicates that this is caused by the very long duration of these measurements. Indeed, if we take into account only a small part of these measurements, then the corresponding plot is identical to the one for the `woolthorpe` dataset. Again, IP addresses above the parabola are caused by per-destination load balancing. Over time, the routes to these destinations evolve, which removes the tendency of the tool to observe them every other round.

## 3 Causes of the observed dynamics

It is acknowledged that load-balancing routers play a significant role in the observed dynamics of routes with `traceroute`-like measurements [4, 15]. Previous work also suggests that routing dynamics play a key role in the continuous discovery of new IP addresses in our measurements [2]. This section identifies the strong role played by these factors in our observations.

These two factors play different roles. Suppose first that there is no load balancing. In this case, a measurement will discover routing changes as they occur, and the longer a measurement lasts, the more IP addresses it will observe (because more changes will occur). If on the contrary there are no routing changes but load balancing is used, then performing more measurement rounds will lead to observe more IP addresses, independently of the time elapsed between consecutive rounds [3]. The observed dynamics is a combination of these factors.

[3] This is of course only true under certain conditions on the number of measurement rounds and the time elapsed between consecutive rounds.

In order to study this rigorously, we use the `woolthorpe` data set and simulate slower measurements by considering only one out of every two rounds. Fig. 3 presents the number of distinct IP addresses observed with both these measurements, as a function of time elapsed since the beginning of the measurement, and the number of measurement rounds performed.

As expected, less IP addresses are observed over time with the slow measurements than with the faster ones. Fig. 3(a) shows that in a given time interval, performing more measurement rounds therefore allows to discover more IP addresses. This confirms that several measurement rounds are needed to discover all existing routes. This is caused by factors such as load balancing. Conversely, Fig. 3(b) shows that the slow measurements discover more IP addresses *at each round* than the faster ones. Therefore if more time elapses between two consecutive rounds, then each round discovers more IP addresses. This indicates that routes evolve with time.

In both cases, the gap between the plots for the slow and faster measurements are significant, which shows that both factors play an important role in the observed dynamics. This is why we propose a model that incorporates load balancing and route dynamics.

## 4 Model

Our purpose here is to propose relevant and simple mechanisms that reproduce the observations made in Section 2. We do not aim at proposing a realistic model, but rather at providing a first and significant step towards understanding the impact of simple mechanisms on the observed dynamics. This model incorporates four ingredients: the routing topology, the routes from the monitor to the destinations in this topology, load balancing, and routing modifications. For modeling each ingredient, we try to make the simplest choice possible, our goal being to obtain a baseline model which makes it possible to investigate the role of each component, and to which future and more realistic models should be compared.

We represent the topology by a random graph obtained with the Erdös-Rényi model [5], which makes no hypothesis on the structure of the graph and is therefore the simplest model possible. Given a graph representing the topology, we assume that the route between the monitor and a destination is a shortest path, which can be obtained by performing a *breadth-first search (BFS)*. In order to simulate load balancing, each node chooses at random the next node on a shortest path to the destination, and we therefore implement a *random BFS*. It generates a shortest-path tree from the monitor to the destinations by considering the neighbors of explored nodes in a random order. These routing trees will therefore be different from one random BFS to the next, even if the underlying graph does not change.

Second, we need to model changes in the routing topology. We use a simple approach based on link rewiring, or *swap*. It consists in choosing uniformly at random two links $(u, v)$ and $(x, y)$ [4] and swap their extremities, *i.e.* replace them by $(u, y)$ and $(x, v)$.

[4] We choose them such that the four nodes are distinct.

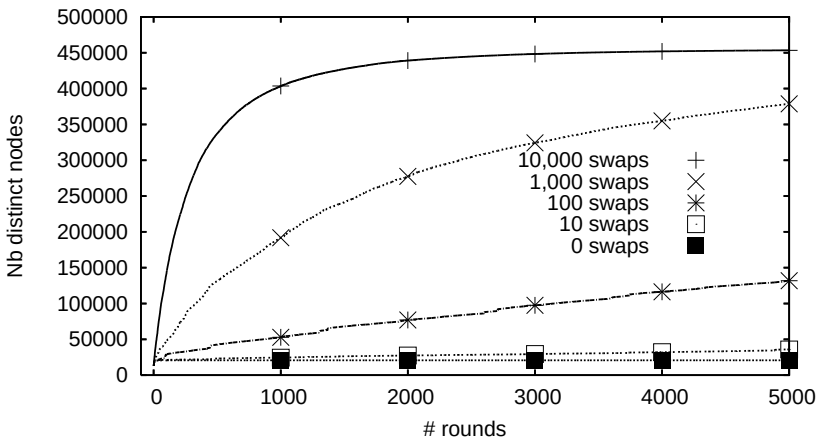

**Fig. 4.** Impact of the number of swaps ($n = 500,000$, $m = 1,000,000$, $d = 3,000$).

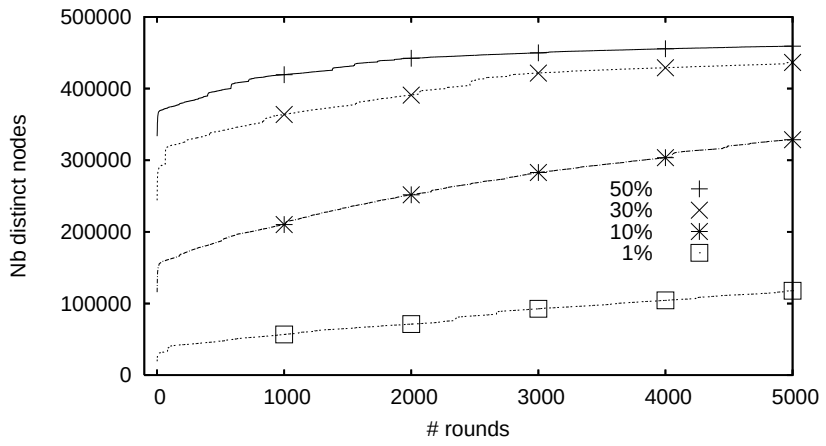

**Fig. 5.** Impact of the number of destinations ($n = 500,000$, $m = 1,000,000$, $s = 50$).

Finally, our simulation setup consists in the following. First, we generate a random graph $G_1$. From $G_1$, we randomly select one node as the monitor and $d$ nodes as the destinations. We then simulate $r$ measurement rounds by iterating the following steps:

1. extract a routing tree $T_i$ from $G_i$ ($i \in [1..r]$) by performing a random BFS from the monitor towards the destinations;
2. modify the graph $G_i$ by performing $s$ random swaps, which produces the graph $G_{i+1}$. $s$ is a parameter of the model.

This process generates a series of $r$ trees $T_1, T_2, \ldots, T_r$ that simulates periodic `tracetree` measurements, on which we can conduct similar analysis as those we performed on real data.

## 5 Results

In this section we show that this model is relevant to explain the dynamics properties presented in Section 2. To that purpose, we perform several simulations varying the parameters of the model: the numbers $n$ of nodes, $m$ of links, $d$ of destinations, and $s$ of swaps per round. Our goals are to find (1) whether the simulations reproduce the observations and (2) how the different parameters impact the results and what are the relations between them.

*Node discovery.* We first study how the simulations behave regarding the evolution of the number of distinct nodes observed over time. Fig. 4 presents such an analysis on a graph with $n = 500,000$, $m = 1,000,000$, $d = 3,000$ and various values of $s$. It shows a similar behavior to the one we observed in real data (see Fig. 1(a)). In particular, for a small number of swaps (less than $100$), the curves present clearly a fast initial growth [5] and then a linear progression. Moreover, the slopes of the curves increase with the number of swaps. This is due to the fact that with a higher number of swaps, the paths to the destinations change even more quickly and thus more nodes are discovered at each step.

The extreme cases are also very informative. For instance, when the underlying graph does not evolve ($s = 0$), there is an initial growth in which all shortest paths

[5] this phase lasts more than 1 round, although it is difficult to visualize it on the plot.

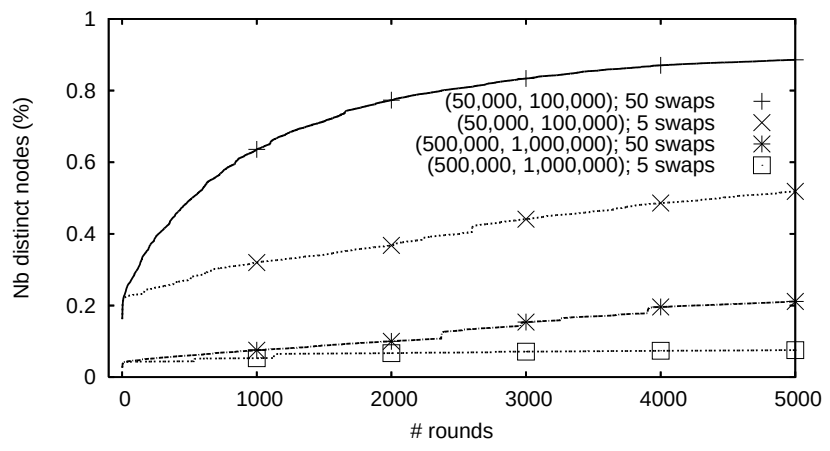

**Fig. 6.** Relation between size and swaps ($m/n = 2$, $d = 3,000$).

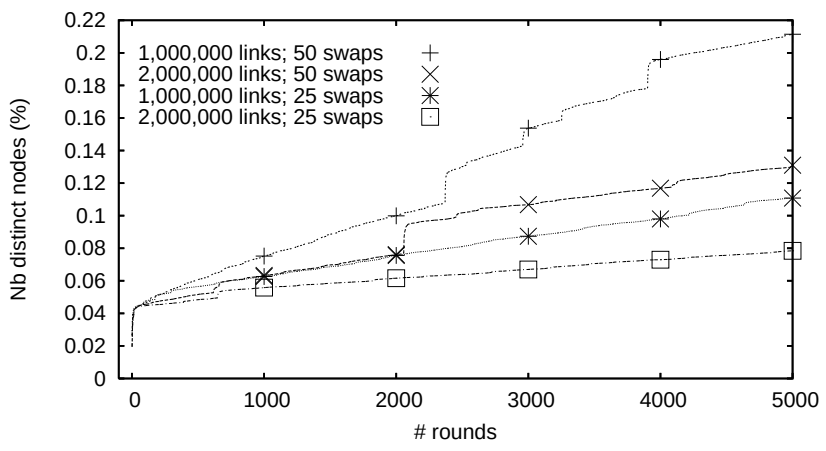

**Fig. 7.** Relation between links and swaps ($n = 500,000$, $d = 3,000$).

are explored. Once all nodes on these paths have been discovered, the curve becomes flat. This confirms that the regular discovery of new IP addresses in real data may stem from route dynamics. At the opposite, when the route dynamics is very high (e.g., $s = 10,000$), almost all the nodes are discovered ($89.49\%$ of the nodes after $2,000$ rounds). Indeed, there are so many changes of the underlying topology – $20,000,000$ swaps compared to $1,000,000$ links – that eventually all nodes will appear at some point on a shortest path [6].

In order to confirm these results, we also study the impact of the number of destinations $d$ on a graph with $n = 500,000$, $m = 1,000,000$ and $s = 50$ (Fig. 5). Intuitively, increasing the number of destinations causes the number of nodes on the shortest paths to the destinations to increase. Indeed, we observe that the initial growth phase, which corresponds to the discovery of all nodes on the shortest paths to the destinations, reaches a higher value when the number of destinations increases. As before, this phase is followed by a linear increase, except for extreme cases in which most of the nodes are discovered. Notice that increasing the number of destinations does not affect much the slope.

*Relations between parameters.* In order to further study the parameter ranges which allow to reproduce the invariants, we now vary several parameters at the same time.

We first set $m/n = 2$ and $d = 3,000$ and then vary $n$ and $s$. Fig. 6 shows that for given values of $n$ and $m$, the value of $s$ does not affect the height of the initial growth. Indeed the number of nodes on shortest paths to the destinations does not depend on the number of swaps. This result confirms that swaps affect mainly the linear part of the curve. In addition, it seems that the slope of the linear part of the curves are similar when the ratio of the number of swaps over the number of links remains the same. The two curves in the middle of Fig. 6 seem to increase with same relative slope, although they have different heights.

We now turn to the relation between the number of swaps and the number of links (Fig. 7). We set $n = 500,000$ and $d = 3,000$. We first observe that, for a given number

[6] note that we never succeed in discovering the entire graph: the Erdös-Rényi model naturally generates nodes with a degree equal to 0, and we choose to keep them in the graph in order to have the same number of nodes in all cases. Since swaps preserve the nodes' degrees, these 0-degree nodes will never be on a path from the monitor to a destination, and will therefore never be discovered.

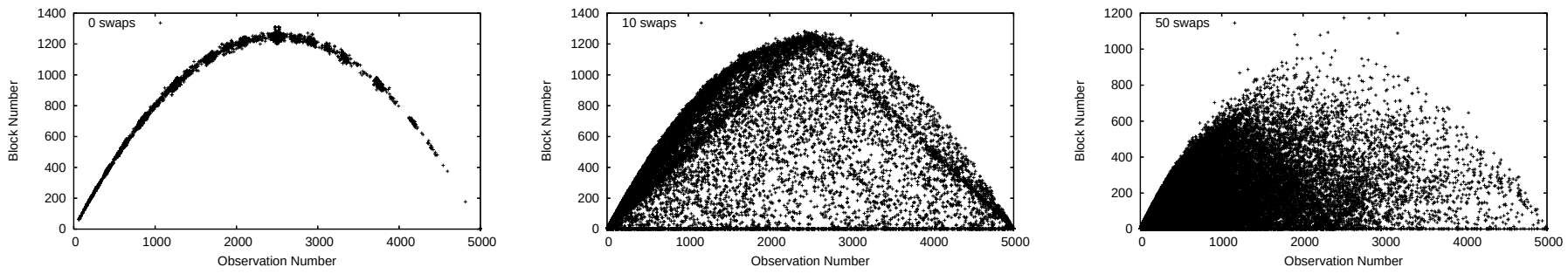

(a) 0 swaps (b) 10 swaps (c) 50 swaps

**Fig. 8.** Observation number vs. block number for various values of $s$ ($n = 500,000$, $m = 1,000,000$, $d = 3,000$).

of swaps, the larger the number of links, the smaller the slope of the corresponding curve. This comes from the fact that, when the number of links increases, a smaller fraction of them is affected by swaps. Second, different curves with a same ratio $s/m$ will have the same slope. We can observe this in the two middle curves. Note that there is a sharp increase in one of them close to $x = 2,000$. It is probably caused by a swap happening very close to the monitor, which therefore affects a larger part of the paths than usual. Yet, this does not affect the slope after such an event.

We performed several simulations with the same parameter values as above, and made similar observations. It is not possible in this case to obtain more rigorous conclusions by, e.g., performing a linear regression on the plots: as we just mentioned, the plots present sharp increases, that greatly impact the slope obtained with a linear regression.

*Observation number vs. block number.* We finally study in Fig. 8 the correlations between the observation number and block number. We fix $n = 500,000$, $m = 1,000,000$ and $d = 3,000$. Again, we performed several simulations with the same parameters and obtained similar results.

For $s = 10$ (Fig. 8(b)), the main invariants we observed in Fig. 1(b) are reproduced: the parabola, the $y = x/2$ line and a dense strip close to the $x$-axis. Notice that, in this case and the others, no triangle appears above the parabola. As already explained in Section 2, this behavior observed on real data comes from an artifact induced by the `tracetree` tool itself in presence of per-destination load balancing. It is therefore natural that the model does not reproduce it.

As already explained in Section 2, the line $y = x/2$ corresponds to nodes that are observed with probability $p = 1/2$ for a given duration, and are not observed before or after. We also observe a high density of nodes on a line with equation $y = (r - x)/2$, $r$ being the total number of rounds performed. This line has a similar explanation: it corresponds to nodes which are observed with probability $p = 1/2$ for a given duration, and are observed *at all rounds* before and after that. Although this line is not present in Fig. 1(b), it sometimes can be observed in other datasets, although not as clearly as here.

When no route dynamics is simulated ($s = 0$, Fig. 8(a)), only the parabola is present, thus confirming that this phenomenon observed in real data is due to load balancing mechanisms which are well captured by the random BFS model. At the opposite, when the number of swaps is too high (Fig. 8(c)), route dynamics get the better of load bal-ancing phenomena and the parabola tends to vanish. The dynamics is so intense that a small number of nodes have a higher block number than expected and are therefore above the parabola.

We showed in Fig. 7 that the number of swaps and links have opposite effects. We studied the correlations between the observation number and the block number for a fixed number of swaps $s$ and different number of links $m$. The analysis confirmed our former statement. In particular, for small values of $m$, the shape of the correlation is degraded as in Figure 8(c). For higher values of $m$, the parabola is more clear as in Figure 8(b) and the plot evolves towards Figure 8(a) as the number $m$ increases, although such an extreme case is impossible to reproduce as soos as $s$ is stricly greater than $0$.

As a conclusion, by exploring the impact of the parameters, we showed that a wide range of their values are relevant. In particular, they all produce a behavior qualitatively similar to what we observed in real data. Note finally that the choice of the Erdös-Rényi model for the underlying topology ensures that such conclusions are not due to the structure of the generated graphs but only to the ingredients we use in the models for load-balancing and routing dynamics.

## 6 Related work

Study of the dynamics of the Internet topology has been tackled both by analyzing the dynamics of individual routes [6–9] and from a more global perspective, mainly at the AS- or IP-level [10–14]. Load balancing has also been acknowledged for playing an important role in the dynamics of routes as measured with traceroute-like tools [15]. Cunha et *al.* [4] used a method for measuring load-balanced routes, i.e. routes containing one or more load-balancing routers, and study their dynamics. Most related to our characterization of the evolution of the Internet topology is the work by Oliveira et *al.* [12], which analyzed the AS topology and shows that real topology dynamics can be modeled as constant-rate birth and death of links and nodes.

Work on modeling the Internet topology and its dynamics can be roughly divided between approaches aiming at reproducing global network characteristics through simple mechanisms, and approaches aiming at realistically mimicking the evolution mechanisms of the topology, e.g. reproducing the criteria taken into account by ASes for creating peering or customer-provider links [16–18]. Tangmunarunkit et *al.* [19] found that network generators based on local properties, such as the degree distributions of nodes, can capture global properties of the topology, such as its hierarchical structure. This shows that simple, non realistic mechanisms such as the ones we introduce may be relevant for capturing important properties.

Whereas most existing works focus on the long-term evolution (i.e. from the Internet birth to current times) of the *physical* AS topology, we are concerned here with the short- to medium-term evolution of the *routing* topology at the IP-level. The routing and physical topology are closely linked but not identical objects. In particular, routing changes can occur in the absence of physical changes. Finally, our model does not take into account node appearance and disappearance, which would be necessary for modeling the long-term topology evolution.

## 7 Conclusion

In this work we conducted periodic measurements of ego-centered views of the Internet topology and studied their dynamics. We isolated invariant characteristics of these dynamics, and identified load balancing and evolution of the routing topology as key factors in the observed properties.

Based on this observation, we proposed a model for the dynamics of the topology, which integrates both load balancing and routing changes. Simulations show that this model captures the main characteristics of the dynamics of the ego-centered views.

Our model is based on simple mechanisms, both for the topology generation and the characteristics of route dynamics. It is therefore not suitable for generating realistic time-evolving topologies. However, the fact that it captures the main characteristics of the observed ego-centered views shows that the factors it mimics play a strong role in the Internet routing topology dynamics, which offers key insight on the understanding of these dynamics. We therefore consider this model as a key step towards the realistic modeling of the Internet topology dynamics, as well as towards its understanding.

Future work lies in two main directions. First, the field of Internet topology modeling is very active, and models far more realistic than random graphs are available. One should explore the combination of our routing mechanisms principles with these topology models, to investigate the role played by the topology structure on the observed dynamics. In particular, our model does not take into account the long term topology evolution, since it does not model node birth or death. Coupling the ingredients of our routing dynamics with, e.g., a growing model for the Internet topology which would reflect its long term dynamics would surely lead to insightful results.

Second, since it is based on random graphs and simple mechanics for load balancing and routing dynamics, our model lends itself well to formal analysis. Obtaining analytical results for the empirical observations we made would be an important step towards quantifying formally the role played by the different factors in the Internet topology dynamics.

**Acknowledgements:** This work is partly funded by the European Commission through the FP7 FIRE project EULER (Grant No.258307). It was also supported in part by a grant from the *Agence Nationale de la Recherche*, with reference ANR-10-JCJC-0202. Finally, we thank Vincent Cohen-Addad, Louis Fournier and Antoine Javelot for their fruitful collaboration, and Matthieu Latapy for enlightening comments.